\documentclass[a4paper]{jpconf}
\usepackage{graphicx}
\usepackage{url}

\def\ha{H$\alpha$}
\def\hb{H$\beta$}

\def\OIII{[O\,\textsc{iii}]}

\begin{document}

\title{HASH: the Hong Kong/AAO/Strasbourg \ha\ planetary nebula database}

\author{
Quentin A  Parker$^{1,2}$, Ivan S Boji\v{c}i\'c$^{1,2}$ and David J  Frew$^{1,2}$
}

\address{$^1$ Department of Physics, The University of Hong Kong, Hong Kong SAR, China}
\address{$^2$ Laboratory for Space Research, The University of Hong Kong, Hong Kong SAR, China}

\ead{quentinp@hku.hk}

\begin{abstract}
By incorporating our major recent discoveries with re-measured and verified contents of existing catalogues we provide,
for the first time, an accessible, reliable, on-line SQL database for essential, up-to date information for all known Galactic planetary nebulae (PNe). 
We have attempted to: i) reliably remove  PN mimics/false ID's that have biased previous  studies and  ii) provide accurate positions, 
sizes, morphologies, multi-wavelength imagery and spectroscopy. We also provide a  link to CDS/Vizier for the archival history of each 
object and other valuable links to external data. With the HASH interface, users can sift, select, browse, collate, investigate, 
download and visualise the entire currently known Galactic PNe diversity. HASH provides the community with the most complete and reliable 
data with which to undertake new science.
\end{abstract}

\section{Introduction}
PNe, the ejected, ionised shrouds of dying stars, are  a complex and brief ($\sim$25,000 year)  phase of late stellar 
evolution.  They offer rich science as vital probes of stellar nucleosynthesis processes in mid to low-mass stars. These stars make-up 90\% of 
all stars above 1 solar mass. They 
provide a detectable, fossil record of stellar mass loss off the AGB/post-AGB and are powerful tracers of our Galaxy's star-forming history. 
Furthermore, they are useful kinematical probes visible to large Galactic distances due to their rich, strong, emission line spectra. These 
scientific levers and others encourage the search for and study of PN in our own and other Galaxies.

We are currently in a golden age of PN discovery. New high sensitivity, wide-field, narrow-band, Galactic plane surveys
undertaken on the UK Schmidt Telescope in Australia \cite{P98}\cite{SHS}, the Isaac Newton telescope on La Palma \cite{Drew05} and now the ESO/VST in 
Chile \cite{Drew14} have facilitated this. These H$\alpha$ surveys have provided significant Galactic
PNe discoveries that have more than doubled the totals accumulated by all telescopes over the previous 260 years, e.g. 
\cite{MASH}\cite{MASHII}\cite{Sabin14}\cite{DSH1} and these proceedings, including posters by Kronberger et al. and Acker et al.

Most new PNe found are more redenned, evolved and of lower surface brightness than previous compilations such as \cite{SEC}\cite{SEC2} and 
\cite{K01} while others are faint but compact and 
more distant. The scope of any future large-scale PNe studies, particularly those of a 
statistical nature or undertaken to understand true PNe diversity and evolution, should now reflect this fresh PN population landscape. Studies 
should make us of the combined sample of $\sim$3500 Galactic PNe now available in our HASH database. HASH takes into account 
recent major discoveries and the power invested in the wide-field, high sensitivity, high resolution, multi-wavelength imaging surveys now available 
across much of the electromagnetic spectrum.

Our group, now based at Hong Kong University, has played a lead role not only in the discovery of over 1300 new Galactic PNe but has also 
made some key contributions in understanding and quantifying the PNe phenomena. These include multi-wavelength identification, 
verification and testing techniques, e.g. \cite{FP10}\cite{P12}, provision of accurate H$\alpha$ and \OIII\, fluxes for large numbers of PNe, 
\cite{FBP13}\cite{F14a}\cite{Kov} and the establishment and verification of a new, statistical distance scale based on a robust incarnation 
of a surface-brightness radius relation \cite{FP16}. We have also catalogued thousands of interesting non PNe and re-assigned 
significant numbers of supposed PNe into other object-types. All these activities, outputs and experience have been combined and 
incorporated into HASH in one form or another.

This paper can provide only a short HASH introduction. It briefly describes some of the power and broad functionality that HASH offers. 
A more comprehensive journal paper detailing the extensive capabilities of HASH is in preparation (Boji\v{c}i\'c et al).

\section{Rationale for establishing the HASH PN research platform}
During our discovery and verification work for the MASH \cite{MASH}\cite{MASHII} and IPHAS \cite{Sabin14} catalogues it was clear that  previous PNe compilations 
were variable in quality and integrity. This is unsurprising as they contain heterogeneous assemblages of PNe identified, mis-identified and re-identified 
again over many decades by dozens of astronomers working with a wide variety of telescopes, detectors, 
resolutions, wave-bands and sensitivities. Furthermore, it became clear that not only does the availability of high-resolution, deep, optical, 
narrowband imaging surveys of high astrometric integrity provide the basis for significant new discoveries, but that these surveys also allow 
us to revisit the identity, morphologies and recorded positions for most PNe in existing catalogues. The advent of other high quality broad-band optical, near-IR, 
mid-IR and radio wide-field surveys provides strong, additional diagnostic and discriminatory power across the Galaxy.
Taken together we can now  construct a new type of PNe repository that can effectively federate all these new data and the 
extant spectroscopy into a single `research platform' for investigating and evaluating all objects currently or previously identified as a PN in our 
Galaxy.

We have largely completed this task and the first HASH release is available in beta-form (contact authors for access). 
We have tried to resolve problems with 
mimics and dubious identifications for a purer, uncontaminated sample. We give more accurate positions, diameters, morphologies and 
designations. Vast amounts of disparate and scattered data have been consolidated and is presented in an intuitive, graphical 
MySQL-based interface for ease of use. Astronomers can use HASH on-line to interrogate, collate, evaluate, visualise and 
download chosen PN multi-wavelength images, spectra or tables from the entire currently known Galactic PNe population. HASH provides 
the community with the most complete, carefully vetted and reliable data to do this. HASH allows for highly efficient and effective examination and 
comparison of a variety of key observational PN properties for very large samples. 

\section{Input Catalogues}



First we ingested data from the three largest catalogues of Galactic PNe: the Strasbourg-ESO catalogue of PNe (SEC 
\cite{SEC}) and its supplement \cite{SEC2}; Version 2000 of the Catalogue of Galactic Planetary Nebulae \cite{K01} and the Macquarie/AAO/
Strasbourg H$\alpha$ (MASH) catalogues \cite{MASH}\cite{MASHII}, together with 159 new PNe from the related IPHAS survey 
in \cite{Sabin14}. Our database contains all true, likely, possible, and misclassified PNe from these major catalogues.    We then incorporated 
$\sim$400 true and candidate PNe from a large number of recent papers, that have been primarily discovered (or confirmed) optically since 
2001.  The most important are \cite{DSH1}\cite{Boumis03}\cite{Boumis06} and \cite{A15}.  We also include 
a significant number of unpublished candidates found by us from on-going examination of  H$\alpha$ survey data, plus sundry other objects.  
Additional candidate PNe recently found in near/mid-IR surveys were also added, chiefly taken from \cite{P12}\cite{Phillips}\cite{Kwok} and \cite{Ramos} 
 while another 320 PN candidates found in the mid-IR at 24\,$\mu$m were added from \cite{Mizuno}\cite{Wachter} and \cite{Gvara}.

Using the Simbad Criteria Query Tool,
we also searched for all objects called PNe, possible PNe, or proto-PNe (Simbad Object codes: PN, PN? and post-AGB, respectively) in the 
literature.  All objects were cross-checked for duplication and identity (also using \cite{PK67}\cite{Kimes} and \cite{Kerber}), noting that the 
classifications of many nebulae have been fluid over time, e.g. \cite{FP10} and that many current Simbad PN or possible PN identifications 
have turned out to be erroneous. For example \cite{Cohen11} showed that in the GLIMPSE zone 45\% of previously known, pre MASH PNe 
are in fact contaminants (mostly compact H\,II regions).

We reference objects to the main catalogues and compilation papers where we found a  PN identification or 
candidate and whose given positional data and other standard identifiers are input to Simbad for initial assessment. These 
should not be confused with the original, discovery papers for individual or small groups of PNe which we did not 
consider in this work.  For  the majority of  objects readers are referred to the main input catalogues  above which contain an extensive 
discovery list for almost all associated PNe, and of course, to the Simbad database. The original data for each object is recorded in the main MySQL 
table and includes: a unique identifier (usual name), PNG  identifier, coordinate (from the original source), a reference to the 
catalogue of origin and the status of the object (see section \ref{sec:pniden} for information of status flags). 

Over 6000 Galactic objects are currently in our working database, including  $\sim$3500 true, likely, and 
possible PNe together with $\sim$1500 mimics of various kinds (e.g.\cite{Boissay12}) including about 50 transitional objects \cite{Suarez} but not $\sim$500 post-AGB stars and related objects \cite{Torun}\cite{VFP15}.  We are
re-investigating mimics in HASH using multi-wavelength discrimination techniques we developed (e.g.\cite{FP10}). There are 
many  unclassified objects that need more investigation before  they can be properly identified (see \cite{FM10} and \cite{DTSer} for
example case-studies).  

\section{Presentation of the data (online interface)}\label{sec:interface}

The HASH PN database and its online interface is currently hosted and maintained on our own server at HKU. In the near future we will mirror this 
server in Europe in order to secure faster service to users outside of Asia and better availability in case of technical problems. There is also an 
agreement to host a HASH front-end at the CDS in France. 
The backbone of the system is a relational database (MySQL) which provides data consistency  and quick, efficient search and manipulation. 
HASH objects are characterised by a unique set of parameters (id, coordinates and PNG designation) and extensive observational data, collected from the literature and astronomical data repositories and mapped to corresponding objects.

The online interface is organised in {\it views}. The main view is the {\it table view} shown in Fig.1 which provides the tabulated data for all objects 
from the selected sample. The table presents the best available data for each observational parameter. The method for selecting the best available 
data differs from parameter to parameter and it is mostly based on the quality of observational parameters.
We provide other available data on each PN's {\it individual page} - see Fig.2 below. In the {\it image views} one can select one or more pre-made total 
intensity or RGB thumbnail images to be presented for the currently selected sample. 

\begin{figure*}
\centering  
\includegraphics[width=14cm]{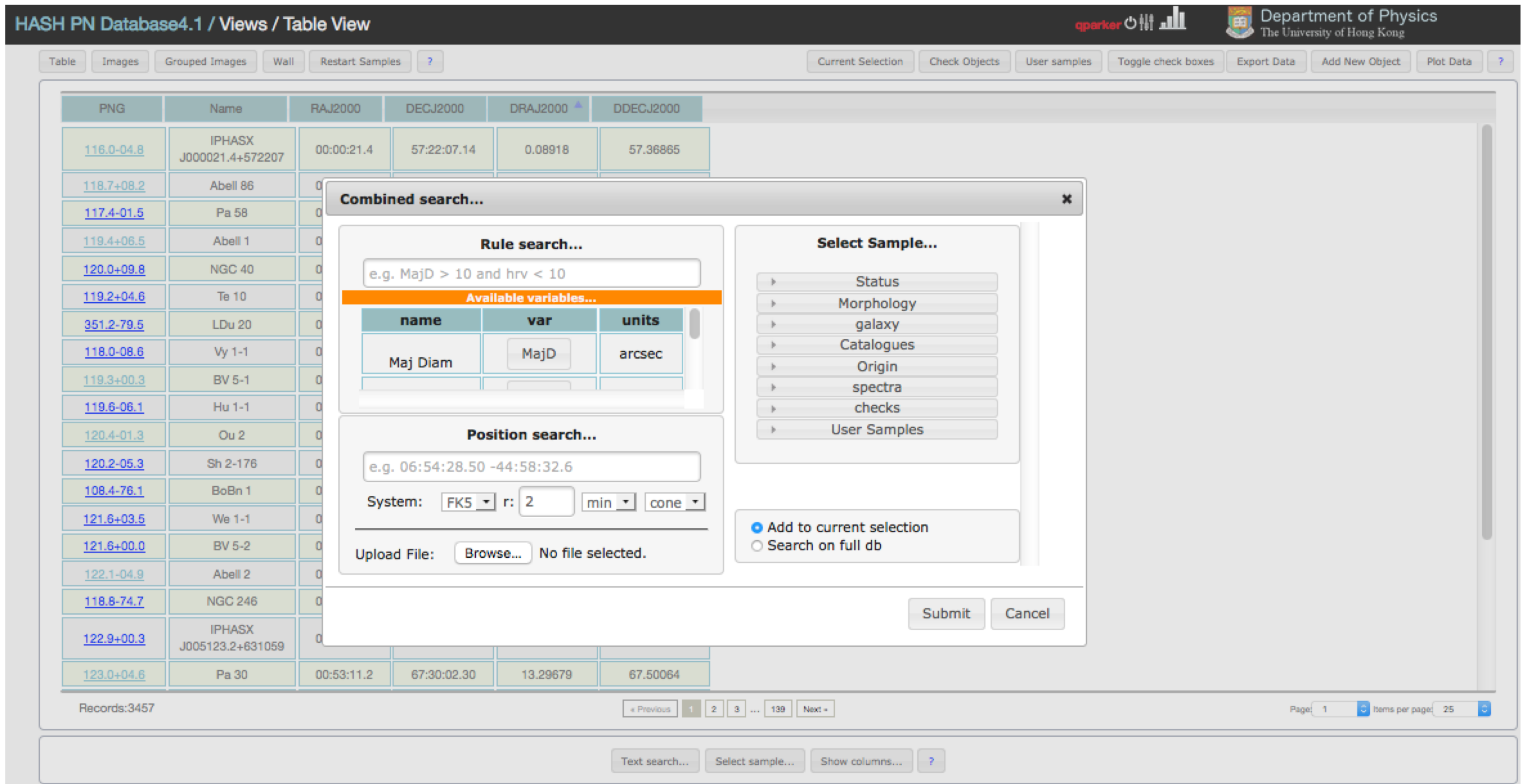}
\caption{ {\small HASH PN basic table view of PN catalogue superimposed with the search box that appears when you want to search by any available 
parameter(s) to select a sub-sample.} }
\label{fig:HASH-table}
\end{figure*}

The search can be on text (e.g. a PN name like NGC~6543) or on one or a combination of  parameters including tabulated 
observational data, positions and association to pre-made samples. In HASH V1.0 we do not include derived data such as PN distances, radial velocities, 
metallicities, etc. These await a future release though much of the data is available to us.

The last sample selected is logged and is the current working sample unless a new sample is selected or the user chooses to restart sample selection. The 
working sample can be subjected to further manipulation (selection of a subsample), exported, saved for a future use  and shared between users. For each 
HASH entry we provide links to the Simbad and Vizier entries based on name or position.  Internet links are also provided to data pages hosted by the Hubble 
Legacy Archive, Planetary Nebula Image Catalogue (PNIC; \cite{PNIC}) and the SPM Kinematic Catalogue of Planetary Nebulae (SPMC; \cite{Lopez}).  Links 
for these appear on the individual HASH pages via clickable logos on the left hand side of the page under the summary data.

\section{HASH multi-wavelength image data}

We have collected and ingested imaging data around each object  from over 30 large scale, multi-wavelength 
surveys from the UV to the radio (as WCS fits files). 
The complete list of surveys incorporated and the full details of all observational parameters available will be published separately 
(Boji\v{c}i\'c et al. in preparation). Particular emphasis is placed on the narrowband \ha\ imaging surveys that have recently become 
available, e.g. \cite{SHS}\cite{Drew05} and \cite{Drew14} that have provided the discovery medium for the significant numbers of new 
MASH and IPHAS PNe in particular.  These modern surveys also provide a pathway to the determination of accurate integrated \ha\ fluxes (e.g. 
\cite{FBP13}\cite{F14a}; see below). Combined RGB colour images for each object from combining various narrow and broad optical, near-IR and 
mid-IR  bands are also provided which also link directly to an associated $RGB$ fits cube that can be downloaded. The field of view, for each image, 
is calculated using the resolution of the survey and angular size of an object. We adopted a simple algorithm based on an intensity percentile to 
present as much information as possible on the object's brightness and structure together with its immediate environment. 

Pre-made composite and/or total intensity images are produced using the APLpy (http://aplpy.github.io/) and Astropy (http://www.astropy.org) packages, the
Montage (http://montage.ipac.caltech.edu/) toolkit and custom made python tools. 
Composite images were made where red 
(R), green (G) and blue (B) colours are for images ordered from the less to more energetic band. Each RGB image layer is scaled 
using the same (linear) scaling function in a way that provides a representative colour image  combination that depends on 
ratios between fluxes in the used bands. 
If the nebular size was unknown or smaller than the survey's PSF we used the
PSF for the aperture size. The measured distribution of data points is used for estimation of the maximum scaling value. The minimum scaling value is 
estimated from the data points outside of the aperture. Prior to RGB composition, constituent fits images were `re-gridded' to a common projection and 
resolution and aligned to have north straight up and east to the left before being combined into fits cubes. 

Each HASH object has an {\it individual page} that contains basic data, links to web pages, image gallery, fits images 
repository, spectral gallery and repository, notes and observational data (refer to on-line documentation for further details). 
The gallery tab provides pre-made images from all available imaging datasets. Overlays of the catalogued position and diameter for each object 
are provided.  An example HASH page for MASH PN PHR J0650+0013 is given in Fig.2. 

\begin{figure*}
\centering  
\includegraphics[width=15cm]{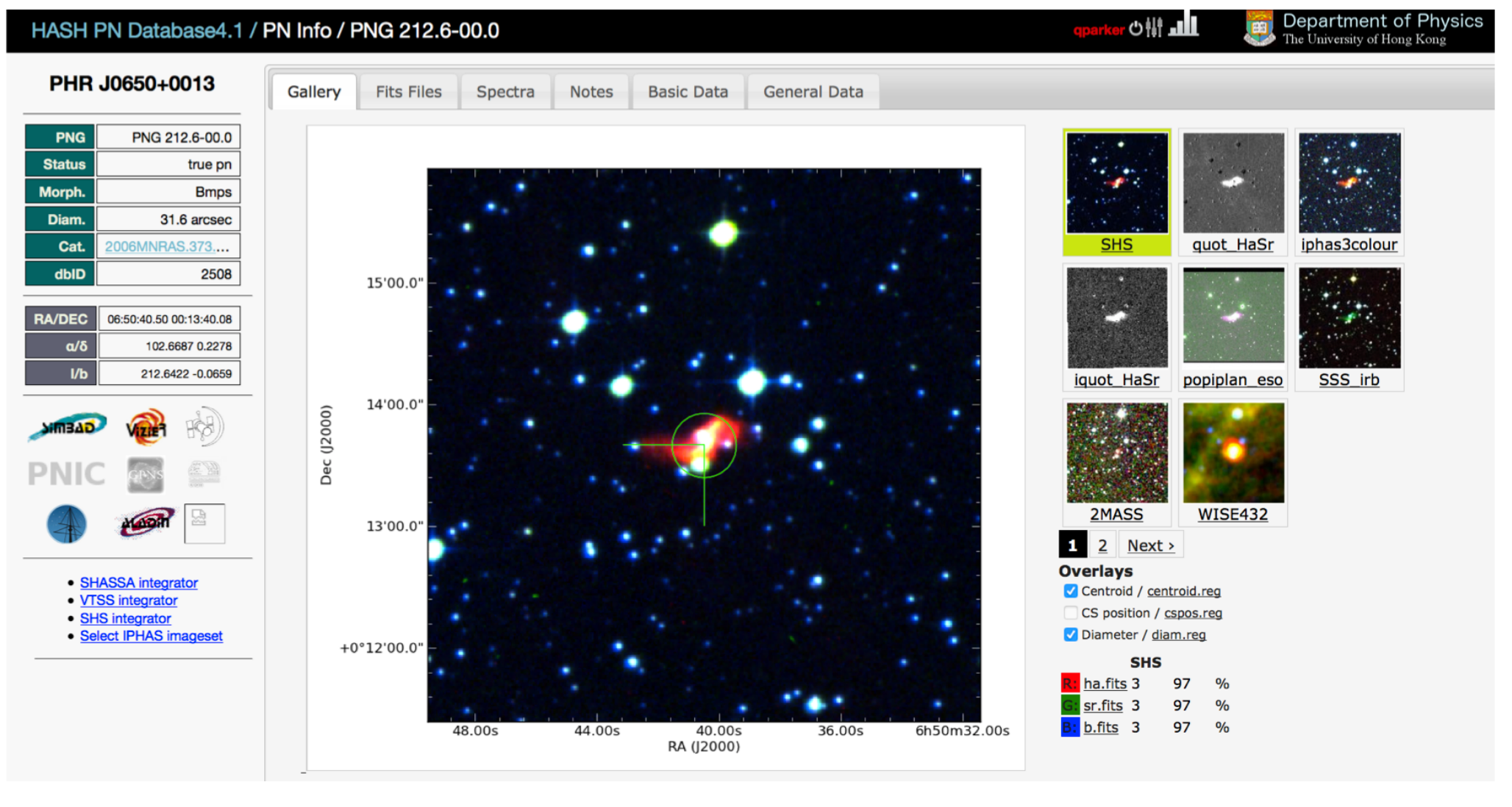}
\caption{ {\small HASH PN  image page for PHR J0650+0013 - a MASH PN.  There is a series of clickable tabs at the top 
of the page which give access to the FITS images for each image, any 1-D spectra, notes pertaining to the object, basic data 
and more general details. All images in thumbnails at the right are clickable.The left hand side column provides summary information for the object 
and direct access via clickable logos to Simbad, Vizier, Aladin, PNIC, HST imagery and other useful sites and date-sets.} }
\label{fig:MASH-PN}
\end{figure*}

\section{Spectroscopic and other data}\label{sec:specdata}
Where possible HASH gives 1-D optical spectra for all PNe.
The primary source is the extensive MASH and IPHAS spectroscopic follow-up programs undertaken by our group and 
collaborators over the last 15 years. These mostly low to intermediate resolution spectra were taken mainly on a series of 2-m telescopes. Another 
major source is the 1050 low-resolution PN spectra from the Stenholm-Acker spectroscopic survey \cite{Sten} that is now available to the 
community for the first time in HASH. Significant numbers ($\sim$1150) of reduced spectroscopic emission line data  for PNe from  \cite{ELCAT}) have 
also been incorporated and are presented as pseudo 1-D spectral  line intensity plots. Other true spectra are also included where available and an 
extensive search of the literature for additional spectra (as described in  \cite{FBP13}) is on-going.  HASH also includes reduced 1-D fits spectra from 
\cite{Boumis03}\cite{Boumis06}\cite{Suarez}\cite{Beaulieu}\cite{Hora99} and \cite{jvds}
 kindly provided by these authors.  Finally, 2-D and 1-D high resolution echelle spectra for accessible PNe  from the San Pedro Martir 
Kinematic Catalogue of Galactic Planetary Nebulae \cite{Lopez} can be accessed from a dedicated clickable logo on the right hand side of the main 
page for each PN when available. Many PNe now have multiple spectra available of different resolutions and wavelength coverage and these are 
made available in HASH. Fig.3 presents a 1-D spectrum for MASH PN PHR J0650+0013 selected by clicking on the spectral tab at the top of 
the individual HASH PN page shown in Fig.2. Common PN emission lines can be selected/de-selected and these are shown as faint, vertical blue 
lines on the plot. If several spectra are available they can be over plotted or selected individually. The cursor can 
also be used to expand the spectra as desired around lines of interest. The spectral files are also available for download for more detailed evaluation 
and measurement.
  
\begin{figure*}
\centering  
\includegraphics[width=14cm]{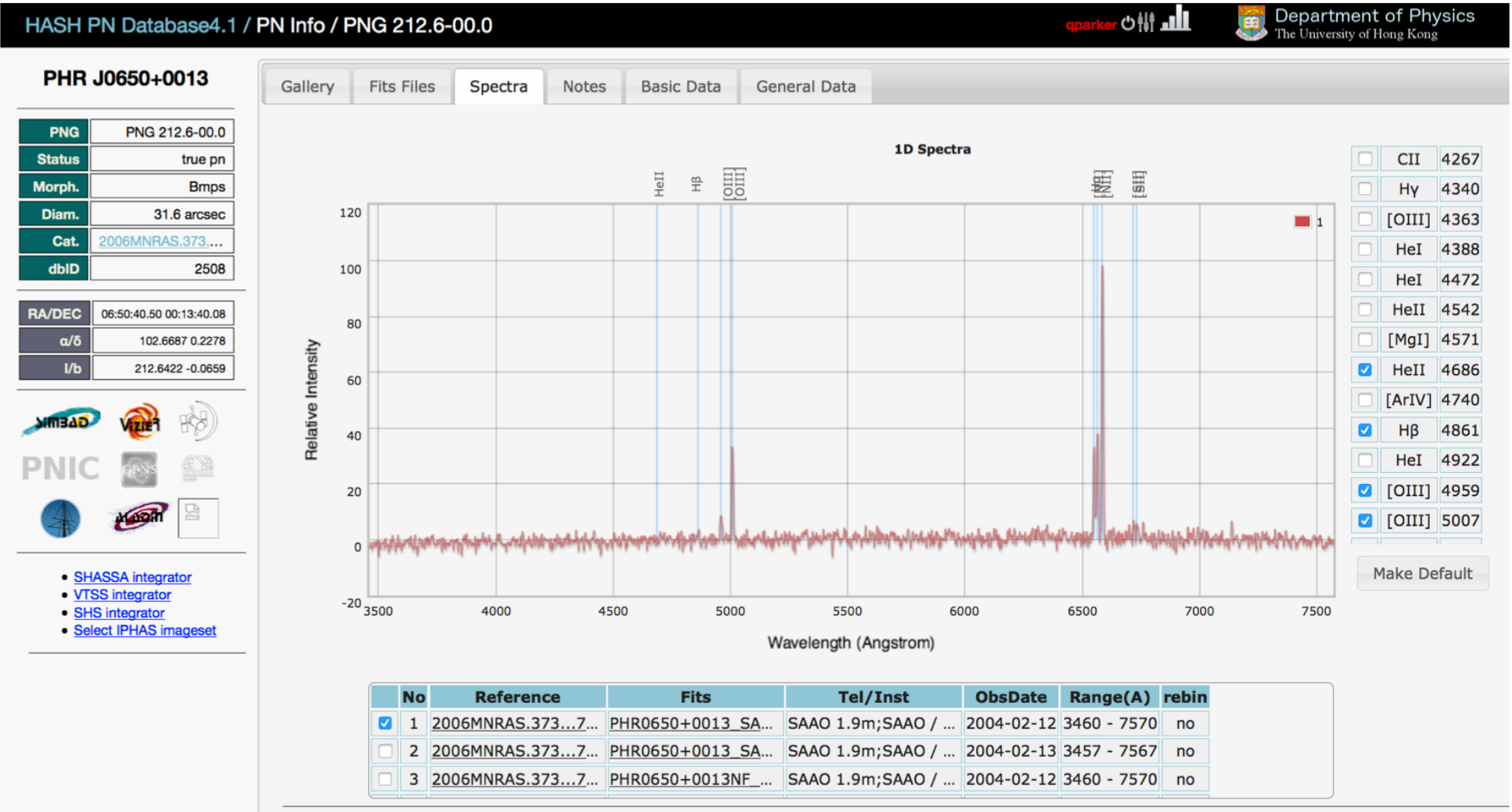}
\caption{ {\small HASH basic spectral page for MASH PN 
PHR J0650+0013 selected by the spectral tab at the top of the individual HASH PN page in Fig.2. 
Common PN emission lines can be selected and are then 
indicated as faint, vertical blue lines on the plot. If several spectra are available they can be over plotted or selected individually. The cursor can 
also be used to expand the spectra as desired.} }
\label{fig:MASH-PN}
\end{figure*}

Besides astrometric data and measurable properties like angular size and more subjective data like morphology we are adding other fundamental 
observational data for each object including integrated \ha, \hb\ and \OIII\ fluxes (e.g. 
\cite{FBP13}\cite{F14a}\cite{Kov} and \cite{A91}), radio flux densities (e.g. \cite{Boj11}), and extinction estimates \cite{SEC}\cite{FP16}\cite{Tylenda}\cite{Giam}
and \cite{SF11}. In future releases radial velocities, emission line ratios, plasma diagnostics, distances and other photometric 
measurements will be included. More extensive central star positions will also be provided as they become available.

\section{The integrity of the HASH database: PN status}\label{sec:pniden}
 
We define a PN following \cite{FP10}. Identification is  complicated by the wide variety of morphologies, 
ionization characteristics and brightness distribution exhibited by PNe. These reflect stages of nebular evolution, progenitor 
mass and chemistry and the possible influence of common envelope binaries, magnetic fields or even sub-solar planets. It is not surprising that 
contamination of previous catalogues has been a problem given such variables and the variable quality of the older data upon which identifications have 
been made. However, we have robustly tested a range of criteria to eliminate contaminants done by assessing morphology, emission-line intensities 
and widths, ionization structure, systemic  velocity, ionized mass and the properties of the central star (where possible) as well as considering 
environment and multi-wavelength imaging properties \cite{Cohen07}. This rigorous process is detailed in \cite{FP10} and we utilised a range of diagnostic diagrams 
presented by \cite{FP10}\cite{FP16}\cite{Sabin13}\cite{F14b}  and Frew, Parker and Boji\v{c}i\'c  (these proceedings).

The HASH PN status flag is adopted from \cite{MASH} and we classify objects  as true ({\bf T}), likely ({\bf L}), possible ({\bf P}) 
and not PN ({\bf N}). As a preliminary step for non MASH/IPHAS PNe in HASH we initially adopt the most recent PN status from the literature. Using 
the available spectroscopic data and the extensive, modern, multi-wavelength imagery now conveniently available for each object in HASH we 
carefully re-examine each object against its current status. On this basis we assign objects as {\bf T} that are confirmed PN with indicative 
multi-wavelength PN-type morphologies, PN spectral features and sometimes presence of an obvious CSPN. {\bf L} indicates an object which is 
likely to be a PN but whose imagery or spectroscopy are not completely conclusive or unavailable. {\bf P} indicates a possible PN where the 
morphology and spectroscopy are insufficiently conclusive, usually due to a combination of low S/N spectra, insufficient wavelength coverage, very 
low surface brightness or indistinct nebulosity. {\bf N} indicates an object which has either been identified as an object of different nature and 
the current data does not provide any additional evidence of a possible PN status, or the newly available combined data that HASH provides strongly 
points to a non PN identification. We believe we have created the purest and most homogeneously assessed catalogue of PN as a result of this work.
\begin{figure*}
\centering  
\includegraphics[width=11cm]{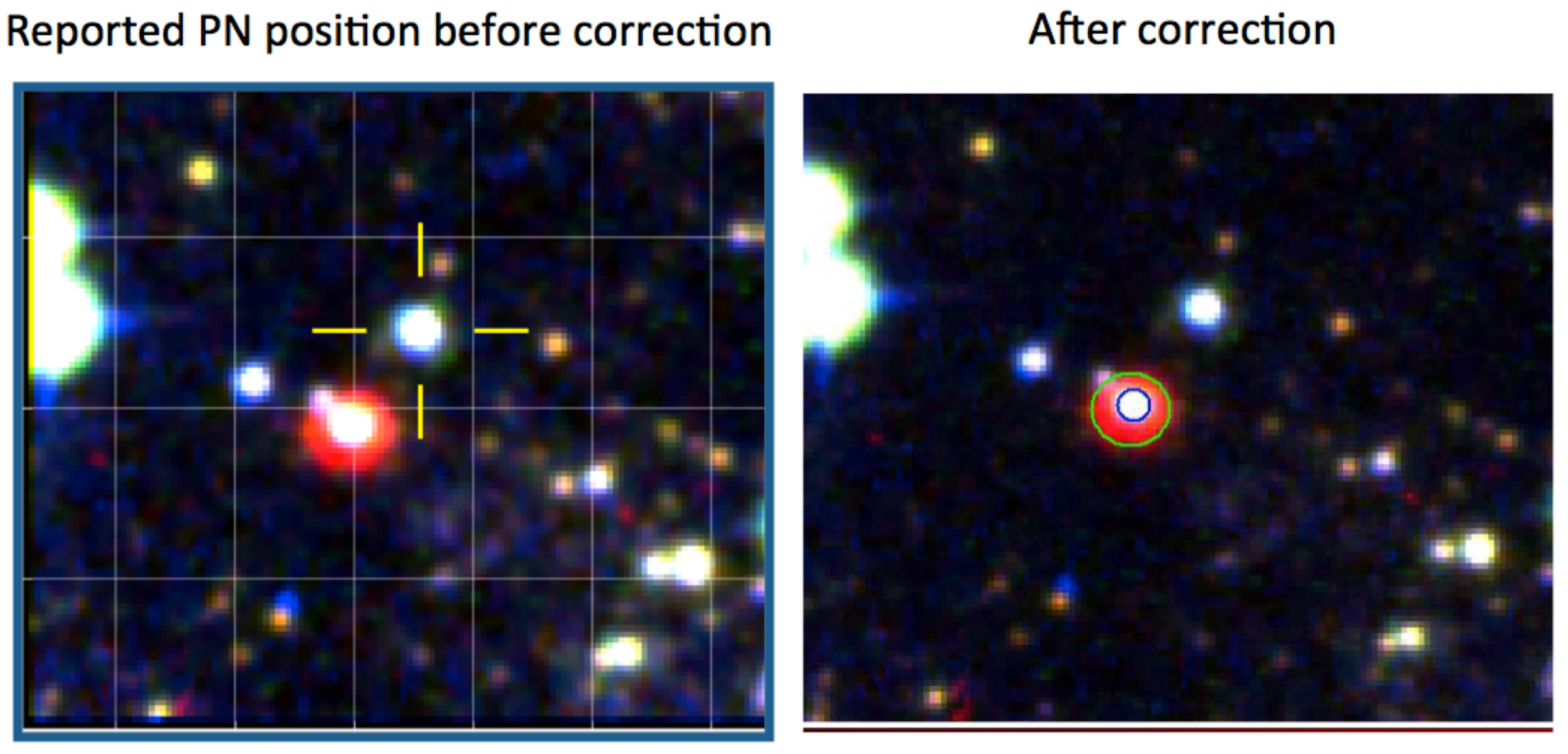}
\caption{
{\small PN PBOZ~29 with position from Simbad based on \cite{Kerber} at left and the new HASH position 
for this PN on the right. The offset is 14.3~arcseconds and the image is 100$\times$100~arcseconds with orientation NE to the top left. The HASH multi-
wavelength combined RGB H-alpha, R-band and B-band images reveal the true PNe and not the star with which this compact PN had been confused in SIMBAD.}
}
\label{fig:PBOZ}
\end{figure*}
Once the status is established we compare the tabulated position against the accurate astrometric grid of the object's best resolution optical image. 
Despite recent papers of `accurate' PN co-ordinates (e.g. \cite{Kimes}, \cite{Kerber}  and even MASH),  it was surprising how 
poor some PN positions remained. Sometimes this is because only part of the PN extent was used to provide the position from broad band data 
whereas on-line surveys in various narrow and broad-bands with accurate astrometry now enables a more robust mechanism to check PNe 
positions.  Sometimes the wrong object has been identified (see Fig.4 for PN PBOZ~29). Where differences are clear a new measurement of the 
object centroid is obtained that then replaces the old value. New estimates of the 
major and minor axes in arcseconds are also made at this time. Accurate, homogeneously derived PNe positions and sizes now form part of  HASH 
and were obtained by aperture and centroid fitting to mainly \ha\ images at the 80\% contour.  Full details of this process will be in the main journal 
paper. However, 29\% of PNe had positional offsets of 10~arcseconds or greater and many had smaller offsets down to 1~arcecond, all of which we 
have now corrected.


\section{Conclusions}\label{sec:conclusions}

We are providing a fully integrated, functional and stable PN research platform that will be fully VO  compliant and offers the community a one-stop shop for 
facilitating PN research. HASH is now available for beta-testing (contact the authors). Future releases will occur  as new data and published results are 
ingested. These data and discoveries will be vetted and checked against our own standards and techniques to ensure, as far as is possible, the on-going 
integrity and homogeneity of the database. In future releases more functionality and parameters are planned. For example we intend to 
incorporate radial velocities and emission line ratios and associated plasma diagnostics for every PN where this is possible in a future release. HASH requires regular curation and maintenance of content, even on an individual PN basis, between major releases and we are committed to this task. Extension to the LMC and SMC is 
also in train.


\end{document}